\documentclass{llncs}
\usepackage{psfig}

\def\aRb#1#2{\hbox{$#1R\,#2$}}
\def\edf{\mbox{\ $\stackrel{\mbox{\tiny def}}{=}$\ }}
\begin{document}

\title{An Application-Level Dependable Technique\\
for Farmer-Worker Parallel Programs}

\author{Vincenzo De Florio, Geert Deconinck, Rudy Lauwereins}

\institute{Katholieke Universiteit Leuven\\
Electrical Engineering Dept - ACCA\\
Kard. Mercierlaan 94 -- B-3001 Heverlee -- Belgium}

\maketitle

\begin{abstract} 
An application-level technique is described for
farmer-wor\-ker parallel applications which allows a worker to be added
or removed from the computing farm at any moment of the run time without
affecting the overall outcome of the computation.  The technique is based on
uncoupling the farmer from the workers by means of a separate module which
asynchronously feeds these latter with new
``units of work'' on an on-demand basis, %
%
%
and on a
special feeding strategy based on bookkeeping the status of each work-unit.
An augmentation of the LINDA model is finally proposed to exploit
the bookkeeping algorithm for tuple management.  
\end{abstract} 
\section{Introduction} 
Parallel computing is nowadays the only technique that can be used in order
to achieve the impressive computing power needed to solve a number of
challenging problems; as such, it is being employed by an ever growing
community of users in spite of what we feel as two main disadvantages,
namely:

\begin{enumerate}
\item harder-to-use programming models, programming techniques and 
development tools---if any,---which sometimes translate into programs
that don't match as efficiently as expected with the underlying
parallel hardware, and
\item the inherently lower level of dependability that characterizes
any such parallel hardware i.e., a higher probability for events like
a node's permanent or temporary failure.
\end{enumerate}

A real, effective exploitation of any given parallel computer
asks for solutions which take into a deep account the above 
outlined problems.

Let us consider for example the synchronous farmer-worker algorithm i.e., a
well-known model for structuring data-parallel applications: a master
process, namely the farmer, feeds a pool of slave processes, called
workers, with some units of work;  then polls them until they return their
partial results which are eventually recollected and saved. Though quite
simple, this scheme may give good results, especially in homogeneous,
dedicated environments. 

But how does this model react to events like a failure of a worker, or more
simply to a worker's performance degradation due e.g., to the exhaustion
of any vital resource?
Without substantial modifications, this scheme is not able to cope with these
events---they would seriously affect the whole application or its overall
performances, regardless the high degree of hardware redundancy 
implicitly available in any parallel system.
The same unflexibility prevents a failed worker to re-enter 
the computing farm once it has regained the proper operational state.

As opposed to this synchronous structuring, it is possible for example to
implement the farmer-worker model by de-coupling the farmer from the workers
by means of an intermediate module, a dispatcher which asynchronously
feeds these latter and supplies them with new units of work on an
on-demand basis.
This strategy guarantees some sort of a dynamic balancing of the workload
even in heterogeneous, distributed environments, thus
exhibiting a higher matching to the parallel hardware. The Live Data Structure
computational paradigm, known from the LINDA context, makes this
particularly easy to set up (see for example~\cite{CaGe1,CaGe2,pvmlinda}).

With this approach it is also possible to add a new worker at run-time
without any notification to both the farmer and the intermediate
module---the newcomer will simply generate additional, non-distinguishable
requests for work.  But again, if a worker fails or its performances
degrade, the whole application may fail or its overall outcome be affected
or seriously delayed.  This is particularly important when one considers
the inherent loss in dependability of any parallel (i.e., replicated)
hardware. 

Next sections introduce and discuss a modification to the above sketched
asynchronous scheme, which inherits the advantages of its parent and offers
new ones, namely:  

\begin{itemize} 
\item it allows a non-solitary,
temporarily slowed down worker to be left out of the processing farm as
long as its performance degradation exists, and 
\item it allows a non-solitary worker which has been permanently affected by
some fault to be definitively removed from the farm, 
\end{itemize} 
both of them without affecting the overall outcome of the computation,
and dynamically spreading the workload among the active processors in a way
that results in an excellent match to various different MIMD
architectures.
\section{The Technique}
For the purpose of describing the technique we define the following
scenario:  a MIMD machine disposes of $n+2$ identical ``nodes'' ($n>0$), or
processing entities, connected by some communication line. On each node a
number of independent sequential processes are executed on a time-sharing
basis.  A message passing library is available for sending and receiving
messages across the communication line. A synchronous communication
approach is used: a sender blocks until the intended receiver gets the
message. A receiver blocks waiting for a message from a specific sender, or
for a message from a number of senders. When a message arrives, the
receiver is awaken and is able to receive that message and to know the 
identity of the sender.  Nodes are numbered from 0 to $n+1$.  Node 0 is
connected to an input line and node $n+1$ is connected to an output line. 

\begin{itemize}
\item Node 0 runs:
\begin{itemize}
\item a Farmer process, connected by the input line to an external
producer device. From now on we consider a camera as the producer device.
A control line wires again the Farmer to the camera, so that this latter can
be commanded to produce new data and eventually send this data across the
input line;
\item a Dispatcher process, yet to be described.
\end{itemize}

\item Node $n+1$ runs a Collector process, to be described later on,
connected by the output line to an external
storage device e.g., a disk;

\item Each of the nodes from 1 to $n$ is purely devoted to the execution 
of one instance of the Worker process. Each Worker is connected
to the Dispatcher and to the Collector processes.
\end{itemize}
\begin{figure}
\psfig{file=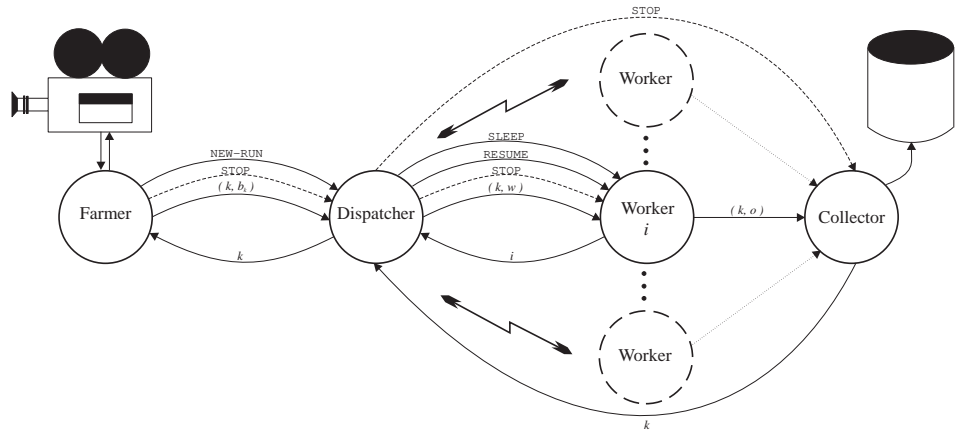,width=4.9in}
\caption{Summary of the interactions among the processes.}\label{f1}
\end{figure}
\subsection{Interactions Between the Farmer and the Dispatcher}
\label{iFS}
On demand of the Farmer process, the camera sends it an input image.
Once it has received an image, the Farmer performs a predefined,
static data decomposition, creating $m$ equally sized sub-images, or blocks.
Blocks are numbered from 1 to $m$, and are represented by variables
$b_i, 1\le i\le m$. 

The Farmer process interacts exclusively with the camera and with the
Dispatcher process. 

\begin{itemize}
\item
Three classes of messages can be sent from the Farmer process to the
Dispatcher (see Fig.~\ref{f1}): 

\begin{enumerate}
\item a {\tt NEW\_RUN} message, which means:
``a new bunch of data is available'';
\item a {\tt STOP} message, which means that no more input
is available so the whole process has to be terminated;
\item a couple $(k,b_k), 1\le k\le m$ i.e., an integer which identifies a particular block
(it will be referred from now on as a ``block-id''), followed by the block
itself.
\end{enumerate}

\item
The only type of message that the Dispatcher process sends to the Farmer
process is a block-id i.e., a single integer in the range $\{1,\dots,m\}$
which expresses the information that a certain block has been fully processed
by a Worker and recollected by the Collector (see~\S\ref{iWC}.)
\end{itemize}

At the other end of the communication line, the Dispatcher is ready to
process a number of events triggered by message arrivals.  For example,
when a class-3 message comes in, the block is stored into a work buffer
as follows:

\begin{tabbing}
tab \= tab \= tab \= tab \= tab \kill
\> {\tt receive} $(k, b_k)$   \\
\> $s_k \leftarrow$ {\tt DISABLED} \\
\> $w_k \leftarrow b_k$      
\end{tabbing}

\noindent
(Here, {\tt receive} is the function for receiving an incoming message,
$\vec{s}$ is a vector of $m$ integers pre-initialized to {\tt DISABLED}, 
which represents some status information that will be described later on, and
$\vec{w}$ is a vector of ``work buffers'', i.e., bunches of memory able to
store any block.
{\tt DISABLED} is an integer which is not in the set $\{1,\dots,m\}$.
The ``$\leftarrow$'' sign is the assignment operator.)

As the Farmer process sends a class-1 message, that is, a {\tt NEW\_RUN}
signal, the Dispatcher processes that event as follows: 

\begin{tabbing}
tab \= tab \= tab \= tab \= tab \kill
\>  $\vec{s} \leftarrow 0$   \\
\>  {\tt broadcast RESUME}
\end{tabbing}

that is, it zeroes each element of $\vec{s}$ and then broadcasts
the {\tt RESUME} message to the whole farm.

\vspace*{2pt}

When the first image arrives to the Farmer
process, it produces a series $(b_i)_{1\le i\le m}$, and then
a sequence of messages $(i,b_i)_{1\le i\le m}$.  Finally, the Farmer
sends a {\tt NEW\_RUN} message. 

Starting from the second image, and while there are images to process
from the camera,
the Farmer performs the image decomposition in advance, thus creating
a complete set of $(k, b_k)$ couples. These couples are then sent to the 
Dispatcher on an on-demand basis: as soon as block-id $i$ comes in, couple
$(i, b_i)$ is sent out. This is done for anticipating the transmission of
the couples belonging to the next run of the computation.
When eventually the last block-id of a certain run has been received,
a complete set of ``brand-new'' blocks is already in the hands of 
the Dispatcher; at that point, sending the one {\tt NEW\_RUN} message
will simultaneously enable all blocks.

\subsection{Interactions Between the Dispatcher and the Workers}
\label{iSW}
The Dispatcher interacts with every instance of the Worker
process.

\begin{itemize}
\item
Four classes of messages can be sent from the Dispatcher to the Workers
 (see Fig.~\ref{f1}):
\begin{enumerate}
\item a {\tt SLEEP} message, which sets the receiver into a wait condition;
\item a {\tt RESUME} message, to get the receiver out of the waiting state;
\item a {\tt STOP} message, which makes the Worker terminate;
\item a $(k,w)$ couple, where $w$ represents the input
data to be elaborated.
\end{enumerate}

\item
Worker $j$, $1\le j\le n$, interacts with the Dispatcher by sending it 
its worker-id message, i.e., the $j$ integer. This happens when
Worker $j$ has finished dealing with a previously sent $w$ working buffer
and is available for a new $(k,w)$ couple to work with.
\end{itemize}

In substance, Worker $j$ continuously repeats the following loop:

\begin{tabbing}
tab \= tab \= tab \= tab \= tab \kill
\> {\tt send $j$ to Dispatcher}                  \\
\> {\tt receive {\em message\/} from Dispatcher} \\
\> {\tt process {\em message\/} }
\end{tabbing}

Clearly, {\tt send} transmits a message.
The last instruction, in dependence with the class
of the incoming message, results in a number of different
operations:

\begin{itemize}
\item if the message is a {\tt SLEEP}, the Worker waits until the arrival 
of a {\tt RESUME} message, which makes it resume the loop, or the arrival
of any other message, which means that an error has occurred;
\item if it is a {\tt STOP} message, the Worker breaks the loop and exits
the farm;
\item if it is a $(k,w)$ couple, the Worker starts computing
the value $f(w)$, where $f$ is some user-defined function e.g.,
an edge detector.
If a {\tt RESUME} event is raised during
the computation of $f$, that computation is immediately abandoned and
the Worker restarts the loop.
Contrarywise, the output couple $(k, f(w))$ is sent to the Collector process.
\end{itemize}

When the Dispatcher gets a $j$ integer from Worker $j$,
its expected response is a new $(k,w)$ couple, or a {\tt SLEEP}.
What rules in this context is the $\vec{s}$ vector---if all entries
of $\vec{s}$ are {\tt DISABLED}, then a {\tt SLEEP} message is sent to Worker $j$.
Otherwise, an entry is selected among those with the minimum non-negative
value, say entry $l$, and a $(l, b_l)$ message is then sent as a response.
$s_l$ is finally incremented by 1.

More formally, considered set
\( S=\{ s \in \vec{s} \,|\, s\neq {\tt DISABLED}\}, \)
if $S$ is non-empty it is possible to partition $S$ according to the
equivalence relation $R$ defined as follows:
\[ \forall (a,b)\in S\times S : \aRb{a}{b} \Leftrightarrow s_a = s_b. \]
So the blocks of the partition are the equivalence classes:
\[ [x] \edf \{ s\in S \,|\, \exists y\in\{1\dots m\} \ni' 
	      (s = s_y) \wedge (s_y = x) \}. \]
Now, first we consider
\[ a = \min\{ b \,|\, \exists b\ge0 \ni' [b]\in \frac{S}{R}\}; \]
then we choose $l\in [a]$ in any way e.g., pseudo-randomly; finally,
message $(l, b_l)$ is sent to Worker $j$, 
$s_l$ is incremented, and the partition is reconfigured accordingly.
If $S$ is the empty set, a {\tt SLEEP} message is generated.

In other words, entry $s_i$ when greater than or equal to 0 represents
some sort of a priority identifier (the lower the value, the higher 
the priority for block $b_i$). The block to be sent to a requesting
Worker process is always selected among those with the highest priority;
after the selection, $s_i$ is updated incrementing its value by 1.
In this way, the content of $s_i$ represents
the degree of ``freshness'' of block $b_i$: it substantially counts
the number of times it has been picked up by a Worker process;
fresher blocks are always preferred. 

As long as there are ``brand-new'' blocks i.e., blocks with a freshness
attribute of 0, these are the blocks which are selected and distributed. 
Note that this means that as long as the above condition is true,
each Worker deals with a different unit of work; on the contrary,
as soon as the last brand-new block is distributed, the model admits 
that a same block may be assigned to more than one Worker.

This is tolerated up to a certain threshold value; if any $s_i$ becomes
greater than that value, an alarm event is raised---too many workers
are dealing with the same input data, which might mean that they are
all affected by the same problem e.g., a software bug resulting in an
error when $b_i$ is being processed. We won't deal with this special
case. Another possibility is that two or more Workers had finished
their work almost at the same time thus bringing rapidly a flag to
the threshold. Waiting for the processing time of one block may
supply the answer.

A value of {\tt DISABLED} for any $s_i$ means that its corresponding block
is not available to be computed. It is simply not considered
during the selection procedure.
\subsection{Interactions Between the Workers and the Collector}
\label{iWC}
Any Worker may send one class of messages to the Collector;
no message is sent from this latter to any Worker
 (see Fig.~\ref{f1}).

The only allowed message is the couple $(k,o)$ in which $o$ is the 
fully processed output of the Worker's activity on the $k^{\hbox{\tiny th}}$ block.

The Collector's task is to fill a number of ``slots'', namely
$p_i, i=1,\dots,m$, with the outputs coming from the Workers.
As two or more Workers are allowed to process a same block thus
producing two or more $(k,o)$ couples, the Collector runs
a vector of status bits which records the status of each slot:
if $f_i$ is {\tt FREE} then $p_i$ is ``empty'' i.e., it has never
been filled in by any output before; if it is {\tt BUSY}, it already
holds an output. $\vec{f}$ is firstly initialized to {\tt FREE}.

For each incoming message from the Worker, the Collector repeats
the following sequence of operations:

\begin{tabbing}
tab \= tab \= tab \= tab \= tab \kill
\> {\tt receive $(k,o)$ from Worker}  \\
\> {\tt if $f_k$ is equal to FREE}    \\
\> \>  {\tt then}                     \\
\> \> \> \>{\tt send $k$ to Dispatcher}\\
\> \> \> \>{\tt $p_k \leftarrow o$}              \\
\> \> \> \>{\tt $f_k \leftarrow $ BUSY}          \\
\> \> \> \>{\tt check-if-full}        \\
\> \>  {\tt else}                     \\
\> \> \> \>{\tt detect}               \\
\> {\tt endif}
\end{tabbing}

\noindent 
where:
\begin{description}
\item[{\tt check-if-full}] checks if, due to the last arrival,
all entries of $\vec{f}$ have become {\tt BUSY}. In that case,
a complete set of partial outputs has been recollected and,
after some user-defined post-processing (for example,
a polygonal approximation of the chains of edges produced by
the Workers), a global output can be saved,
and the flag vector re-initialized:

\begin{tabbing}
tab \= tab \= tab \= tab \= tab \kill
\> {\tt if $\vec{f}$ is equal to BUSY}    \\
\> \>  {\tt then}                     \\
\> \> \> \>{\tt post-process $\vec{p}$}   \\
\> \> \> \>{\tt save $\vec{p}$}           \\
\> \> \> \>{\tt $\vec{f}\leftarrow$ FREE} \\
\> {\tt endif}
\end{tabbing}

\item[{\tt detect}] is a user-defined functionality---he/she may choose
to compare the two $o$'s so to be able to detect any inconsistency
and start some recovery action, or may simply ignore the whole message.
\end{description}

Note also that an acknowledgment message (the block-id) is sent 
from the Collector
to the Dispatcher, to inform it that an output slot has been occupied i.e.,
a partial output has been gathered. This also means that the Farmer
can anticipate the transmission of a block which belongs to the next
run, if any.
\subsection{Interactions Between the Collector and the Dispatcher}
\label{iCD}
As just stated, upon acceptance of an output, the collector
sends a block-id, say integer $k$, to the Dispatcher---it is the only message
that goes from the Collector to the Dispatcher.

The Dispatcher then simply acts as follows:
\begin{tabbing}
tab \= tab \= tab \= tab \= tab \kill
\>{\tt $s_k \leftarrow$ DISABLED} \\
\>{\tt send $k$ to Farmer}
\end{tabbing}

\noindent
that is, the Dispatcher ``disables'' the $k^{\hbox{\tiny th}}$ unit of work---set
$S$ as defined in~\S\ref{iSW} is reduced by one element and consequently
partition $\frac SR$ changes its shape; then the block-id is propagated
to the Farmer (see Fig.~\ref{f1}).

On the opposite direction, there is only one message that may travel from the
Dispatcher to the Collector: the {\tt STOP} message that means that no more
input is available and so processing is over. Upon reception of this
message, the Collector stops itself, like it does any other receiver in the farm.
\section{Discussions and Conclusions}
The just proposed technique uses asynchronicity in order to
efficiently match to a huge class of parallel architectures.
It also uses the redundancy which is inherent to parallelism
to make an application able to cope with events like e.g.,
a failure of a node, or a node being slowed down, temporarily or not.

\begin{itemize}
\item If a node fails while it is processing block $k$, then
no output block will be transferred to the Collector. When no more
``brand-new'' blocks are available, block $k$ will be assigned to
one or more Worker processes, up to a certain limit. During this phase
the replicated processing modules of the parallel machine may be thought of
as part of a hardware redundancy fault tolerant mechanism.
This phase is over when any Worker module delivers its output to the
Collector and consequently all others are possibly
explicitly forced to resume their processing loop or, if too late,
their output is discarded;
\item if a node has been for some reason drastically slowed down,
then its block will be probably assigned to other possibly non-slowed
Workers. Again, the first who succeeds, its output is collected;
the others are stopped or ignored.
\end{itemize}

In any case, from the point of view of the Farmer process, all these
events are completely masked. The mechanism may be provided to a user in the
form of some set of basic functions, making all technicalities concerning
both parallel programming and fault tolerance transparent to the programmer.

Of course, nothing prevents the concurrent use of other fault tolerance
mechanisms in any of the involved processes e.g., using watchdog timers
to understand that a Worker has failed and consequently reset the proper
entry of vector $\vec{f}$. The ability to re-enter the farm may also 
be exploited committing a reboot of a failed node and restarting 
the Worker process on that node.

\subsection{Reliability Analysis}

In order to compare the original, synchronous farmer-worker model with the
one described in this paper, a first step is given by observing that the
synchronous model depicts a {\em series system\/} 
\cite{Johnson} i.e., a system in which each element is required
not to have failed for the whole system to operate. This is not the case of
the model described in this paper, in which a subset of the elements,
namely the Worker farm, is a
{\em parallel system\/} \cite{Johnson}: if at least one Worker has not
failed, so it is for the whole farm subsystem. Note how Fig.~\ref{f1} may be
also thought of as the reliability block diagram of this system.

\begin{figure}
\input{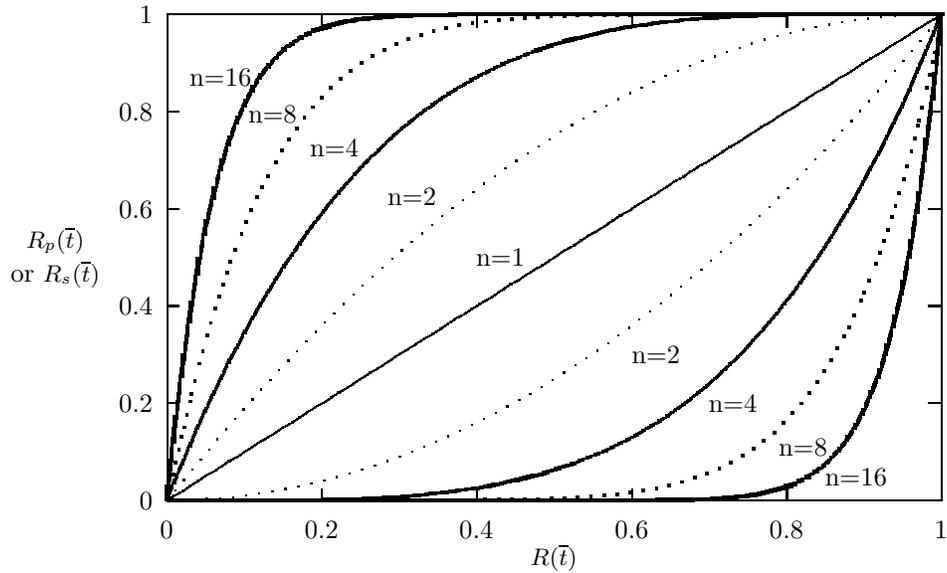}
\caption{For a fixed value $\overline t$, a number of graphs of 
$R_p(\overline t)$ (the reliability of the parallel system)
and $R_s(\overline t)$ (the reliability of the series system)
are portrayed as functions of $R(\overline t)$, the reliability of a Worker
at time $\overline t$, and $n$, the number of the components.
Each graph is labeled with its value of $n$; those above the diagonal
portray reliabilities of parallel systems, while those below the diagonal
pertain series systems.
Note that for $n=1$ the models coincide, while for any $n>1$ 
$R_p(\overline t)$ is
always above $R_s(\overline t)$ except when $R(\overline t)=0$ 
(no reliable Worker) and when
$R(\overline t)=1$ (totally reliable, failure-free Worker).}\label{f2}
\end{figure}

Considering the sole farm subsystem, if we let $C_i(t), 1\le i\le n$ be the
event that Worker on node $i$ has not failed at time $t$, and we let
$R(t)$ be the reliability of any Worker at time $t$ then, under the
assumption of mutual independency between the events, we can conclude that:

\begin{equation}
R_s(t) \edf P( \bigcap_{i=1}^{n} C_i(t)) = \prod_{i=1}^{n} R(t) = (R(t))^n
\end{equation}

\noindent
being $R_s(t)$ the reliability of the farm as a series system, and

\begin{equation}
R_p(t) \edf 1 - P( \bigcap_{i=1}^{n} \overline{C_i}(t)) =
            1 - \prod_{i=1}^{n} (1 - R(t)) = 1 - (1-R(t))^n
\end{equation}

\noindent
where $R_p(t)$ represents the reliability of the farm as a parallel system.
Of course failures must be independent, so again data-induced errors are
not considered. Figure~\ref{f2} shows the reliability of the farm in a series
and in a parallel system as a Worker's reliability goes from 0 to 1.

\subsection{An Augmented LINDA Model}
The whole idea pictured in this paper may be implemented in
a LINDA tuple space manager (see for example \cite{CaGe1,CaGe2}).
Apart from the standard functions to access ``common'' tuples,
a new set of functions may be supplied which deal with ``book-kept tuples''
i.e., tuples that are distributed to requestors by means of
the algorithm sketched in~\S\ref{iSW}. As an example:

\begin{description}
\item[{\tt fout}] (for fault tolerant {\tt out}) may create a book-kept tuple 
i.e., a content-addressable object with book-kept accesses;
\item[{\tt frd}] (fault tolerant {\tt rd}) may get a copy of a matching 
book-kept tuple, chosen according to the algorithm in~\S\ref{iSW};
\item[{\tt fin}] (fault tolerant {\tt in}) may read-and-erase a matching 
book-kept tuple, chosen according to the algorithm in~\S\ref{iSW},
\end{description}
\noindent
and so on. The ensuing augmented LINDA model results in an abstract, 
elegant, efficient, dependable, and transparent mechanism to exploit a 
parallel hardware. 

\subsection{Future Directions}
The described technique is currently being implemented on a
Parsytec CC system with the EPX/AIX environment \cite{CCpg}
using PowerPVM/EPX \cite{powerpvm}, a homogeneous version
of the PVM message passing library;
it will also be tested in heterogeneous, networked environments 
managed by PVM. Some work towards the definition and the development
of an augmented LINDA model is currently being done.
\paragraph{Acknowledgments.}
This project is partly sponsored by the Belgian Interuniversity Pole of
Attraction IUAP-50, by an NFWO Krediet aan Navorsers, and by the
Esprit-IV project 21012 EFTOS. Vincenzo De Florio is on leave
from Tecnopolis CSATA Novus Ortus.
Geert Deconinck has a grant from the 
Flemish Institute for the Promotion of Scientific and Technological Research
in Industry (IWT). Rudy Lauwereins is a Senior Research Associate of the
Belgian National Fund for Scientific Research.
%
%

%
\end{document}